\documentclass[prb,showpacs,amsmath,amssymb,notitlepage,superscriptaddress]{revtex4-1}
\usepackage{graphicx}
\usepackage{bm}
\usepackage{bbm}
\usepackage{subfigure}
\usepackage{amssymb}
\usepackage{amsmath}
\usepackage{xcolor}
\usepackage{xcolor,xspace,colortbl,rotating}
\usepackage{amsfonts}
\usepackage{tabulary}
\usepackage{color}

\bibstyle{unsrt}

\begin{document}
\title[Spin caloric effects in antiferromagnets]{Spin caloric effects in antiferromagnets assisted by an external spin current}
\date{\today}

\author{O. Gomonay}
\address{Institute f\"ur Physik, Johannes Gutenberg Universit\"at Mainz, D-55128 Mainz, Germany}
\address{National Technical University of Ukraine KPI, 03056, Kyiv, Ukraine}
\author{Kei Yamamoto}
\address{Department of Physics and Astronomy, The University of Alabama, Alabama 35487, USA \\
	and Center for Materials for Information Technology (MINT), The University of Alabama, Alabama 35401, USA}
\address{Institute f\"ur Physik, Johannes Gutenberg Universit\"at Mainz, D-55128 Mainz, Germany}
\author{Jairo Sinova}
\address{Institute f\"ur Physik, Johannes Gutenberg Universit\"at Mainz, D-55128 Mainz, Germany}
\address{Institute of Physics ASCR, v.v.i., Cukrovarnicka 10, 16253 Praha 6, Czech Republic}
\vspace{10pt}
\begin{abstract}
Searching for novel spin caloric effects in antiferromagnets we study the properties of thermally activated magnons in the presence of an external spin current and temperature gradient. We predict the spin Peltier effect -- generation of a heat flux by spin accumulation -- in an antiferromagnetic insulator with cubic or uniaxial magnetic symmetry. This effect is related with spin-current induced splitting of the relaxation times of the magnons with opposite spin direction.  We show that the Peltier effect can trigger antiferromagnetic domain wall motion with a force whose value grows with the temperature of a sample. At a temperature, larger than the energy of the low-frequency magnons, this force is much larger than the force caused by direct spin transfer between the spin current and the domain wall. We also demonstrate that the external spin current can induce the magnon spin Seebeck effect. The corresponding Seebeck coefficient is controlled by the current density. These spin-current assisted caloric effects open new ways for the  manipulation of the magnetic states in antiferromagnets.  
	\end{abstract}
\maketitle

\section*{Introduction}
Spin caloritronics is a new emerging field which merges spintronics and caloritronics. 
Its focus is on manipulating and controlling magnons and spins with a temperature gradient. Within this field many thermoelectric effects such as Seebeck, Peltier, Joule-Thompson effects were extended to include spin transport phenomena in ferromagnets \cite{Uchida2008, Flipse2012,Kovalev2012,  Flipse2014, Basso2016, Daimon2016,Rezende2016a,Nakata2017a}. Very recently some of the spin caloric effects have been also reported in antiferromagnets \cite{Seki2015,Rezende2016, Wu2016, Holanda2017, Ritzmann2017}. Antiferromagnets are 
promising materials for spintronics and spin-caloritronics because they operate at  higher frequencies compared to ferromagnets and have more magnetic degrees of freedom which could be involved in magnon transport.  However, possible physical mechanisms of the combined thermal and magnon transport phenomena in antiferromagnets still require thorough understanding. In addition, because the magnon dynamics in antiferromagnets and ferromagnets are quite different, new functionalities of antiferromagnetic materials should follow as we explore their behaviour under action of the magnetic fields, electrical currents, and temperature gradients.

The crucial difference between the spin caloric effects in ferro- and antiferromagnets steams from the fact that antiferromagnets posses zero or vanishingly small net magnetization. This entails spin-neutrality of an antiferromagnetic magnon gas and hinders observation of spin caloric effects  in antiferromagnets. To overcome this challenge and to remove spin-degeneracy of the nonpolarized magnon gas one can apply an external magnetic field simultaneously with the temperature gradient, as it was done in several experiments \cite{Seki2015, Wu2016, Holanda2017}. Here we consider another option to stimulate the spin caloric effects in antiferromagnets focusing on the thermo-magnonic effects assisted by an \emph{external spin current}. 

In contrast to the external magnetic field which affects the magnon frequency, the external spin current (or spin accumulation), via spin transfer torque mechanism, can split the degeneracy of the magnon relaxation times (damping coefficients) \cite{Gomonay2013a}. The relaxation times, according to the fluctuation-dissipation theorem (FDT), are the important constituents of the Onsager coefficients which couple the fluxes of magnons, magnon spins, and heat with the conjugated thermodynamic forces (magnon/spin accumulation and temperature gradient). Thus, manipulation of the relaxation times with spin current enables to control the values of the Onsager coefficients and induce the corresponding thermomagnonic effects. 

In the present paper we analyse the magnon spectra of an antiferromagnet and calculate the magnon (spin) Seebeck coefficients as a function of the density and polarization of the external spin polarized current. Basing on the properties of the magnon spectra we predict the spin Peltier effect and describe the magnon (spin) Seebeck effect supported by the spin current. We show that the Peltier and the magnon (spin) Seebeck effects can induce motion of the antiferromagnetic domain walls. We compare the driving forces induced by the thermal and non-thermal mechanisms in the presence of the external spin-polarized current. Combination of the predicted caloritronic and spintronic effects can be used for the effective switching between the different antiferromagnetic states in experiment-friendly geometry.

\begin{figure}
	\centering
	\includegraphics[width=0.95\linewidth]{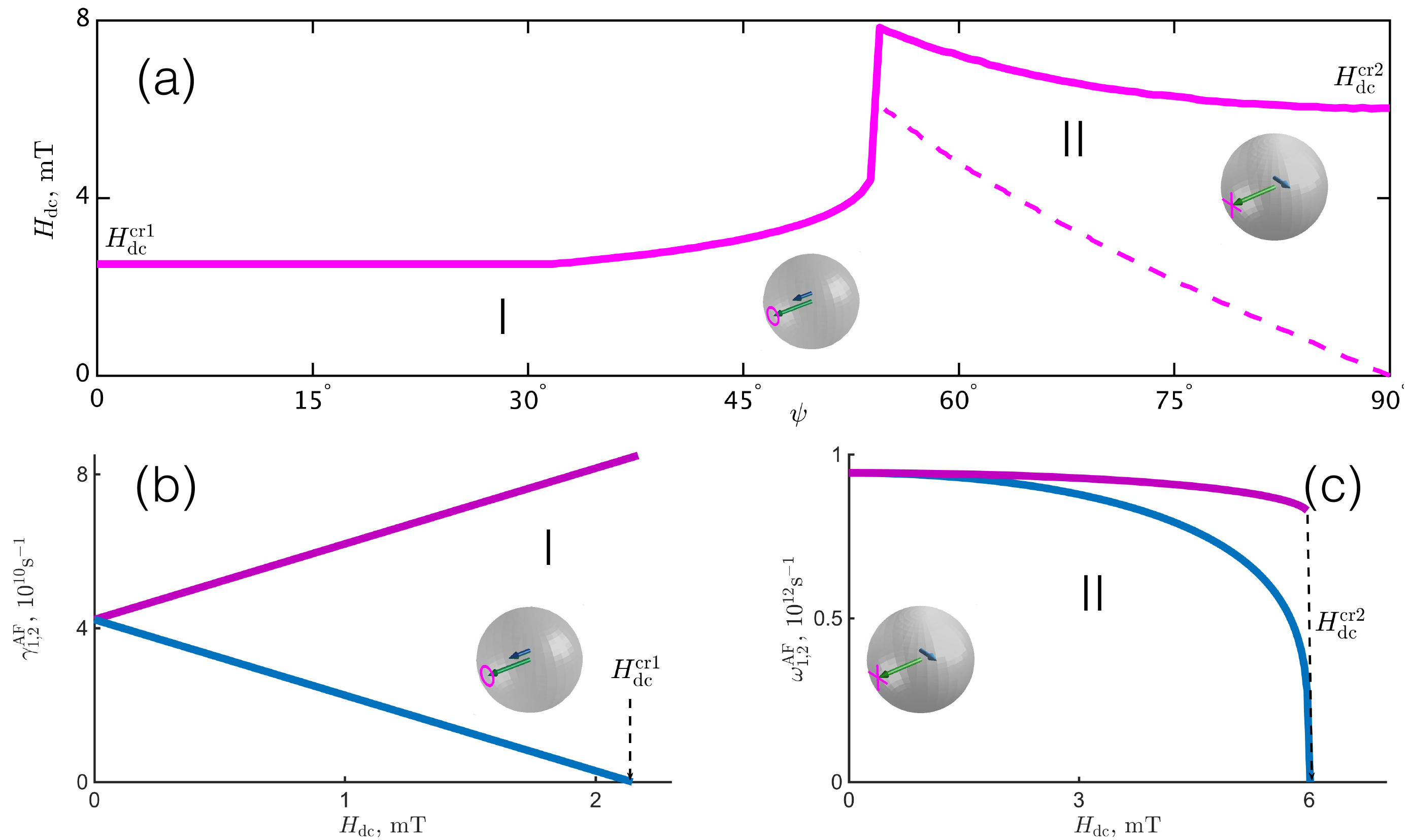}
	\caption{(Color online) (a) Phase diagram of a cubic antiferromagnet in the presence of the external spin current. The value of the critical current $H_\mathrm{dc}^\mathrm{cr}$ (solid line) is calculated from the instability condition $\gamma^\mathrm{AF}_{2}(H_\mathrm{dc}^\mathrm{cr})\cdot\omega^\mathrm{AF}_{2}(H_\mathrm{dc}^\mathrm{cr})=0$ for a fixed $\psi$ (=angle between $\mathbf{s}$ and $\mathbf{n}^{(0)}$ while $\mathbf{s}$ rotates within $xy$ plane). Dotted line separates the states with nondegenerated damping coefficients (area I, circular polarized modes) and nondegenerate frequencies (area II, linearly polarized modes). 
		Panels (b) and (c) show nontrivial current dependencies of the (b) damping coefficients for $\mathbf{s}\|\mathbf{n}^{(0)}$ ($\psi=0^\circ$) and (c) frequencies for $\mathbf{s}\perp\mathbf{n}^{(0)}$ ($\psi=90^\circ$) of the eigen modes. The lines on the spheres show the trajectories of the N\'eel vector for the corresponding eigen modes (the amplitudes are exaggerated). $H_\mathrm{dc}^\mathrm{cr1}\equiv H_\mathrm{dc}^\mathrm{cr}(0^\circ)$, $H_\mathrm{dc}^\mathrm{cr2}\equiv H_\mathrm{dc}^\mathrm{cr}(90^\circ)$.
	}
	\label{fig_phase_diagram}
\end{figure}

\begin{figure}
	\centering
	\includegraphics[width=0.95\linewidth]{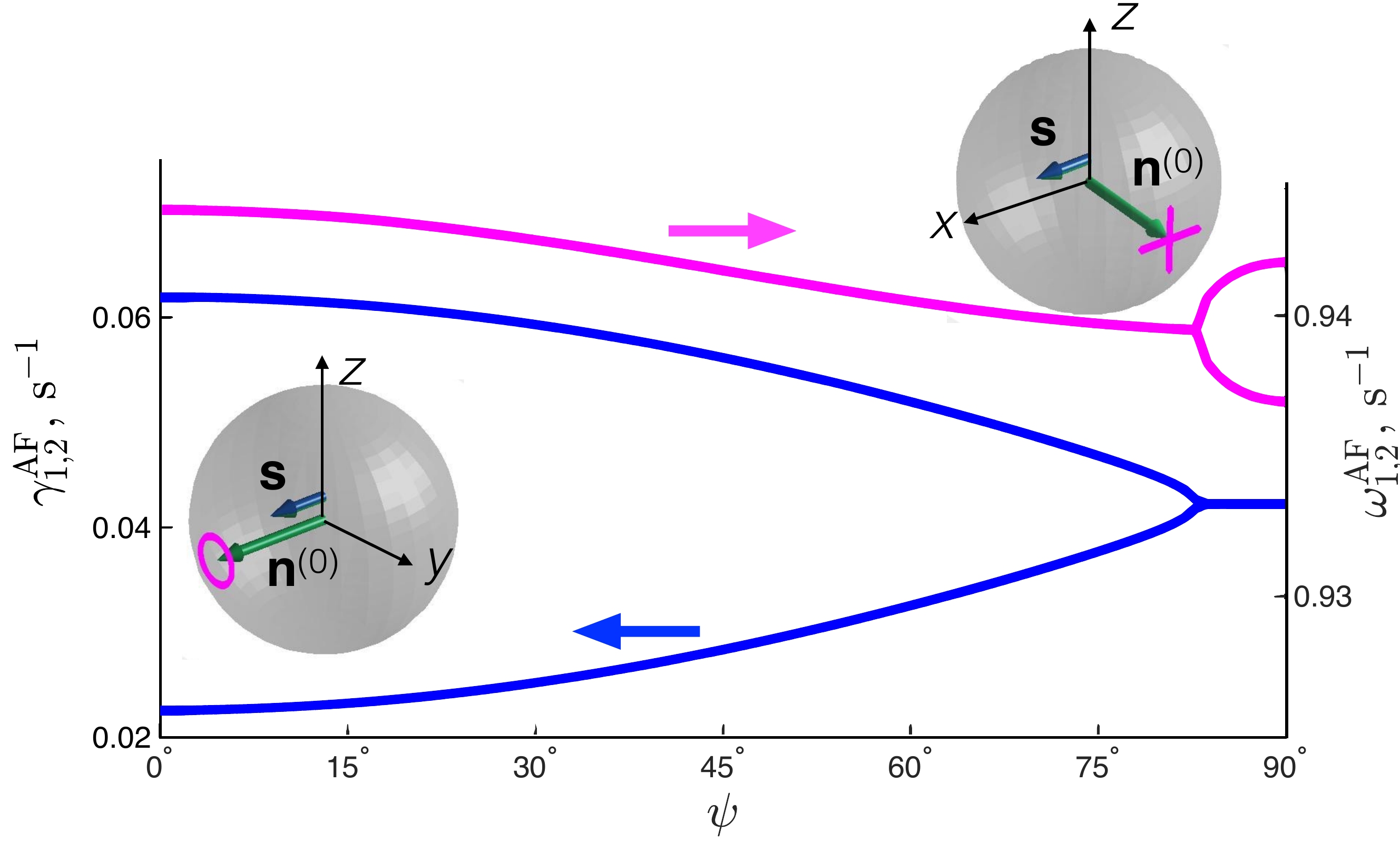}
	\caption{(Color online) Angular dependence of the frequencies (right axis) and damping coefficients (left axis) of the eigen modes. Insets show mutual orientation of $\mathbf{s}$ and $\mathbf{n}^{(0)}$ (blue/short and green/long arrows) and the trajectories (magenta line) of the N\'eel vector for the corresponding eigen modes.  $H_\mathrm{dc}$=1~mT.
	}
	\label{fig_spin_waves_different_types}
\end{figure}

\begin{figure}
	\centering
	\includegraphics[width=0.95\linewidth]{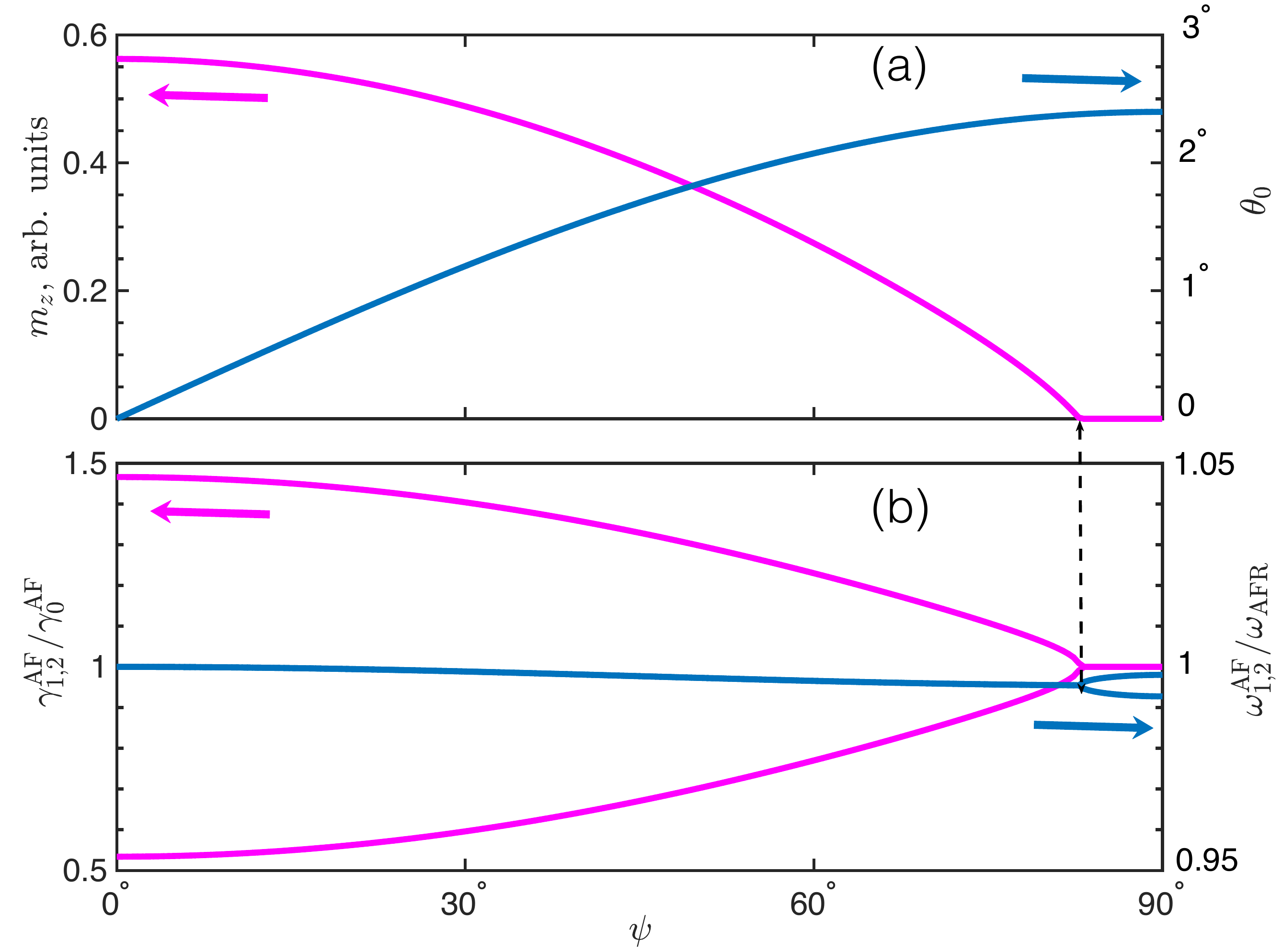}
	\caption{(Color online) Angular dependence of the (a) magnetization (left axis), $\theta_0$ (right axis), and (b) relative damping coefficients (left axis) and relative frequencies (right axes). Double arrow shows the transition point from the circular (area I in Fig.~\ref{fig_phase_diagram}) to the linearly polarized (area II in Fig.~\ref{fig_phase_diagram}) modes.}
	\label{fig_magnetization_relative_damping}
\end{figure}

\section{Model}
We consider a simple case of a collinear antiferromagnet with two magnetic sublattices  $\mathbf{M}_1$ and $\mathbf{M}_2$ which are antiparallel and fully compensate each other in the equilibrium state. While moving, these vectors are slightly tilted forming nonzero magnetization $\mathbf{m}=\mathbf{M}_1+\mathbf{M}_2$.
However, due to the strong exchange coupling between the magnetic sublattices (parametrized with the constant $H_\mathrm{ex}$), the magnetization $\mathbf{m}$ is small and the state of an antiferromagnet is fully described by the  N\'eel vector $\mathbf{n}=\mathbf{M}_1-\mathbf{M}_2$ whose magnitude $|\mathbf{n}|=2M_s$ is fixed well below the N\'eel temperature. In general case the N\'eel vector is space- and time-dependent variable, $\mathbf{n}(\mathbf{r},t)$.

The dissipative dynamics of the N\'eel vector at a finite temperature are driven by i) the anti-damping-like torque \cite{Gomonay2008} $\propto \mathbf{n}\times\mathbf{s}\times\mathbf{n}$, which emerges from the dc spin polarized current 
with spin polarization $\mathbf{s}$, $|\mathbf{s}|=1$, and ii) the field-like torque $\mathbf{n}\times\mathbf{h}_\mathbf{n}$ created by the thermal stochastic field $\mathbf{h}_\mathbf{n}$ \cite{Gomonay2013a, Kim2015d}. 
Equation of motion for the N\'eel vector in the presence of these torques is \cite{Gomonay2008, Cheng2014c}:
\begin{eqnarray}\label{eq_motion_antiferromagnet_initial}
&&\mathbf{n}\times(\ddot{\mathbf{n}}-c^2\nabla^2\mathbf{n}+2\alpha_G\gamma H_\mathrm{ex}\dot{\mathbf{n}}-2\gamma^2H_\mathrm{ex}M_s\mathbf{H}_\mathbf{n})\\
&&=\gamma^2 H_\mathrm{ex}\mathbf{n}\times(H_\mathrm{dc}\mathbf{s}\times\mathbf{n}+2M_s\mathbf{h}_\mathbf{n}),\nonumber
\end{eqnarray}
where $c$ is the magnon velocity, $\alpha_G$ is the Gilbert damping constant, $\gamma$ is the gyromagnetic ratio, and $\mathbf{H}_\mathbf{n}=-\partial w_\mathrm{an}/\partial \mathbf{n}$ is the internal effective field defined by the profile of the magnetic anisotropy energy $w_\mathrm{an}$ (per unit volume). For convenience, we characterise the effective density of spin-polarized current with the value $H_\mathrm{dc}$ which has dimensionality of the magnetic field. In the particular case when the spin current is generated by the spin Hall effect, $H_\mathrm{dc}=\hbar \varepsilon \theta_\mathrm{H}j_\mathrm{dc}/(2ed_\mathrm{AF}M_s)$, where $\hbar$ is the Planck constant, $d_\mathrm{AF}$ is the thickness of a film, $0<\varepsilon\le1$ is the spin-polarization efficiency, $\theta_{H}$ is the bulk Hall angle, $e$ is an electron charge, and $j_\mathrm{dc}$ is the  dc current density in a heavy metal electrode. 

The thermal noise $\mathbf{h}_{\mathbf{n}}$  is characterised by zero average and the correlator
\begin{equation}\label{eq_stochastic}
\langle h_{\mathbf{n},j}(\mathbf{r},t)h_{\mathbf{n},k}(\mathbf{r},t)\rangle=2D\delta_{jk}\delta(t-t^\prime)\delta(\mathbf{r}-\mathbf{r}^\prime),
\end{equation}
where the constant $D$ is related with the damping coefficient and temperature  according to the FDT, as will be discussed below. 

\section{Magnons in the presence of the external spin current}
To calculate magnon transport in different conditions we first analyse the eigen modes of an antiferromagnet in the presence of a spin-polarized current. For this we solve Eq.~(\ref{eq_motion_antiferromagnet_initial}) for the different values of spin current $H_\mathrm{dc}$ and different orientations of $\mathbf{s}$, assuming $\mathbf{h}_\mathbf{n}=0$. For definiteness we consider a cubic antiferromagnet (exemplified by an archetypal antiferromagnetic insulator KNiF$_3$) with the magnetic anisotropy energy (per unit volume)
\begin{equation}\label{eq-anisotropy}
w_\mathrm{an}=-\frac{H_\mathrm{an}^{\perp}}{8M_s^3}(n_x^4+n_y^4+n_z^4).
\end{equation}
Here $H_\mathrm{an}^{\perp}>0$ is the magnetic anisotropy field.

An equilibrium orientation of the N\'eel vector $\mathbf{n}^{(0)}$ is defined from balance of the torques created by the spin current and the internal effective field:
\begin{equation}\label{eq_eqilibrium}
\mathbf{n}^{(0)}\times\left[H_\mathrm{dc}\mathbf{s}\times\mathbf{n}^{(0)}+2M_s\mathbf{H}_\mathbf{n}(\mathbf{n}^{(0)})\right]=0.
\end{equation}
In absence of spin current the N\'eel vector $\mathbf{n}^{(0)}(H_\mathrm{dc}=0)$ points to one of three equivalent easy directions $x$, $y$ or $z$.

Equations for small excitations $\delta\mathbf{n}=\mathbf{n}-\mathbf{n}^{(0)}$ are then obtained by linearising Eq.~(\ref{eq_motion_antiferromagnet_initial}) as follows;
\begin{equation}\label{eq-linear_excitations}
\mathbf{n}^{(0)}\times\left[\delta\ddot{\mathbf{n}}-c^2\nabla^2\delta\mathbf{n}+\gamma H_\mathrm{ex}\underline{(2\alpha_G\delta\dot{\mathbf{n}}-\gamma H_\mathrm{dc}\mathbf{s}\times\delta\mathbf{n})}+\omega^2_\mathrm{AFR}\hat{\Lambda}\delta\mathbf{n}\right]=0,
\end{equation}
where $\omega_\mathrm{AFR}=\gamma\sqrt{H_\mathrm{ex}H_\mathrm{an}^{\perp}}$ is the frequency of antiferromagnetic resonance, $\Lambda_{jk}\equiv (M_s/H_\mathrm{an}^{\perp})\left.(\partial^2 w_\mathrm{an}/\partial n_j\partial n_k)\right|_0$ is the dimensionless matrix whose structure depends on the orientation of $\mathbf{n}^{(0)}$.

The magnon spectrum of the antiferromagnet consists of two degenerate branches which, in absence of spin current, correspond to the circularly-polarized modes with the amplitudes $a_{1,2}(k)$,  frequencies
$\omega^\mathrm{AF}_{1,2}=\sqrt{\omega^2_\mathrm{AFR}+c^2k^2}$, where $\mathbf{k}$ is the wave vector \footnote{~Here, for simplicity, we consider a long-wave limit with $k\ll k_\mathrm{Brill}$ where $k_\mathrm{Brill}$ defines the Brilluen zone boundary. Generalization for the full range of $\mathbf{k}$ is straigtforward once the crystall structure of an antiferromagnet is specified.}, and the relaxation constant (damping  coefficient),  $\gamma^\mathrm{AF}_{1,2}=\gamma^\mathrm{AF}_0\equiv\alpha_G\gamma H_\mathrm{ex}$.
 As oscillations of the N\'eel vector create a dynamic magnetization $\mathbf{m}=\mathbf{n}\times\dot{\mathbf{n}}/(2\gamma H_\mathrm{ex}M_s)$ \cite{Baryakhtar1980}, each of the circularly polarized modes carries a nonzero magnetization ($\propto$ spin) whose direction is parallel/antiparallel to equilibirum orientation of the N\'eel vector:
\begin{equation}\label{eq_mode_magnetization}
	\mathbf{m}_{1,2}=\pm \frac{1}{4\gamma H_\mathrm{ex}M_s^2}\omega^\mathrm{AF}_{1,2}|a_{1,2}(k)|^2\mathbf{n}^{(0)} .
\end{equation}

The properties of the magnon spectra for different $\mathbf{s}$ and $H_\mathrm{dc}$ are  
summarized in Figs.~\ref{fig_phase_diagram}, \ref{fig_spin_waves_different_types}, and \ref{fig_magnetization_relative_damping} plotted for KNiF$_3$ which has the following parameters: $H_\mathrm{ex}$=1200~T, 
$H_\mathrm{an}$=24~mT,  
$\alpha_G=2\cdot10^{-4}$, and $T_N$=246~K \cite{Yamaguchi1999}. Field-current conversion  corresponding to spin-pumping via spin Hall effect with the Hall angle $\theta_\mathrm{H}=0.1$ \cite{Sinova2015} into the sample with thickness $d_\mathrm{AF}=1$~nm can be estimated as $H_\mathrm{dc}/j_\mathrm{dc}=1$ mT/(MA/cm$^2$). We start discussion of these figures from analysis of two limiting cases, $\mathbf{s}\|\mathbf{n}^{(0)}$ and $\mathbf{s}\perp\mathbf{n}^{(0)}$, which represent two qualitatively different scenarios.

\emph{Spin current polarized along the N\'eel vector}, $\mathbf{s}\|\mathbf{n}^{(0)}$, induces current-depending splitting of the relaxation constants of the modes (Fig.\ref{fig_phase_diagram}(b), and data at $\psi=0$ in Figs.~\ref{fig_spin_waves_different_types} and \ref{fig_magnetization_relative_damping}): 
\begin{equation}\label{eq_damping_parallel}
\gamma^\mathrm{AF}_{1,2}=\gamma H_\mathrm{ex}\left(\alpha_G\pm\gamma H_\mathrm{dc}/\omega^\mathrm{AF}_{1,2}\right),
\end{equation}
 leaving the frequencies $\omega^\mathrm{AF}_{1,2}$, polarization, and the magnetizations of the modes unaffected. The mechanism responsible for modification of the relaxation constants 
is based on spin transfer from the external spin current to the magnon mode. In particular, the spin current suppresses the damping constant of the mode with $\mathbf{m}\uparrow\uparrow\mathbf{s}$ and enhances the damping constant of the mode with  $\mathbf{m}\uparrow\downarrow\mathbf{s}$.

The difference between $\gamma^\mathrm{AF}_{1}$ and $\gamma^\mathrm{AF}_{2}$ grows with the current value, $H_\mathrm{dc}$, until one of these constants (depending on the current direction) turns to zero. At this point, $H_\mathrm{dc}=H^\mathrm{cr1}_\mathrm{dc}\equiv\alpha_G\omega_\mathrm{AFR}/\gamma$, the system becomes unstable with respect to coherent rotation of the N\'eel vector \cite{Gomonay2008, Cheng2016}, corresponding value of the spin current sets the phase boundary, as shown in Fig.~\ref{fig_phase_diagram}. 

\emph{Spin current polarized perpendicular to the N\'eel vector}, $\mathbf{s}\perp\mathbf{n}^{(0)}$, modifies the frequencies, polarization and magnetization value of the modes but has no effect on the damping constants (Fig.~\ref{fig_phase_diagram} (c), and data at $\psi=90^\circ$ in Figs.~\ref{fig_spin_waves_different_types} and \ref{fig_magnetization_relative_damping}). In this geometry the spin-polarized current creates a torque which induces a small angle deviation $\theta_0$ of the equilibrium N\'eel vector from the easy axis, as shown in Fig.~\ref{fig_magnetization_relative_damping} (a).  As a result, the eigen modes in this tilted state are linearly polarized (inset in Fig.~\ref{fig_spin_waves_different_types}) and posses zero average magnetization (Fig.~\ref{fig_magnetization_relative_damping} (a)). The corresponding frequencies are nondegenerate and their splitting depends on the current value (Fig.~\ref{fig_phase_diagram}(c)). The state becomes unstable when one of the frequencies goes to zero, which happens at $H_\mathrm{dc}=H^\mathrm{cr2}_\mathrm{dc}\equiv H^\perp_\mathrm{an}/4$.

The described scenarios demonstrate that the external spin current can either modify the relaxation times (when $\mathbf{s}\|\mathbf{n}^{(0)}$) or frequencies (when $\mathbf{s}\perp\mathbf{n}^{(0)}$) of the magnon modes. In the latter case the effect of spin current is similar to the effect of the external magnetic field and we call this scenario field-like, to distinguish it from the other, the antidamping-like one. In a generic case $0<\psi<90^\circ$, the effect of spin current on spectra also follows one of the scenarios, depending on the density and polarization of spin current. This can be seen from Fig.~\ref{fig_spin_waves_different_types} which shows angular dependencies of the damping constants and frequencies. Splitting between the relaxation constants $\gamma_1^\mathrm{AF}-\gamma_2^\mathrm{AF}$ diminishes with deflection of spin polarization from easy axis. The set of $(\psi, H_\mathrm{dc})$ points, in which this difference disappears\footnote{~It is worth to mention that both splittings of the damping coefficient and frequencies disappear at the same $\psi$ value.}, $\gamma_1^\mathrm{AF}-\gamma_2^\mathrm{AF}=0$ (dotted line in Fig.~\ref{fig_phase_diagram}), separates the regions with antidamping-like (area I) and field-like (area II) behavior. These point also corresponds to disappearence of the dynamic magnetization of the modes (Fig.~\ref{fig_magnetization_relative_damping}(a)).  

 To compare the field-like and antidamping-like effects of the external spin current we calculate the relative splitting of the damping coefficients and frequencies normalized to the value at zero current (Fig.~\ref{fig_magnetization_relative_damping}(b)). Maximal frequency splitting is almost two order of values smaller that the splitting of the damping coefficients and thus can be neglected. In what follows we concentrate on the caloritronic effects which appear due to the splitting of the damping coefficients.

\section{Spin Peltier effect in antiferromagnets}
The spin Peltier effect is defined as generation of a magnon heat flux by a spin current through the interface between the heavy metal and the magnetic insulator \cite{Flipse2014}. Up to now the spin Peltier effect was observed in ferro(ferri)magnets \cite{Flipse2012} as cooling or heating of the magnon gas  depending on mutual orientation of the spin current polarization and magnetic moment of a ferro(ferri)magnet. The underlying mechanism is related with  transfer of spin and energy from the heavy metal electrons to the magnons in magnetic insulator via spin transfer or spin-orbit torque. Formally this means that the damping coefficient of the magnon modes which, according to FDT, defines the effective temperature of the magnon gas, increases or decreases due to the external spin current \cite{Apalkov2005}. 

We argue that the same mechanism should provide existence of the spin Peltier effect in antiferromagnetic insulators. To prove this, we solve  Eq.~(\ref{eq_motion_antiferromagnet_initial}) in the presence of thermal noise. From FDT, in assumption that $\hbar\omega_\mathrm{AFR}\ll k_\mathrm{B}T$, it follows that in absence of current the noise correlator (Eq.~(\ref{eq_stochastic})) is equal to
\begin{equation}\label{eq_diffusivity}
D=\frac{2\pi\gamma^\mathrm{AF}k_\mathrm{B}T}{\gamma^2H_\mathrm{ex}M_s},
\end{equation}
where $T$ is the temperature, $k_\mathrm{B}$ is the Boltzmann constant.
However, in configuration, in which the external spin current splits the damping coefficients $\gamma_{1,2}^\mathrm{AF}$ (area I in Fig.~\ref{fig_phase_diagram}(a)), each of the magnon modes acquires different effective temperatures \footnote{~For the case $\mathbf{s}\|\mathbf{n}^{(0)}$ this result was obtained in Ref.\cite{Gomonay2013a}.}:
\begin{equation}\label{eq_effective temperature}
T_{1,2}^\mathrm{eff}=\frac{T\gamma^\mathrm{AF}(H_\mathrm{dc}=0)}{\gamma_{1,2}^\mathrm{AF}}\approx\frac{T}{1\pm H_\mathrm{dc}\cos\psi/H^\mathrm{cr}_\mathrm{dc}(\psi)}.
\end{equation}
The second equality is valid at a relatively small current value, $H_\mathrm{dc}< H^\mathrm{cr}_\mathrm{dc}$, $H^\mathrm{cr}_\mathrm{dc}(\psi)$ dependence is shown in Fig.~\ref{fig_phase_diagram}(a).

As a result, the temperature of the magnon gas, $T^\mathrm{mag}$, calculated as an average magnon energy, depends on the angle $\psi$ between spin current polarization and the N\'eel vector as follows:
\begin{equation}\label{eq_magnon_gas}
\Delta T\equiv T^\mathrm{mag}-T=\frac{H^2_\mathrm{dc}\cos^2\psi}{(H^\mathrm{cr}_\mathrm{dc})^2-H^2_\mathrm{dc}\cos^2\psi}T\ge 0.
\end{equation}

\begin{figure}
	\centering
	\includegraphics[width=0.7\linewidth]{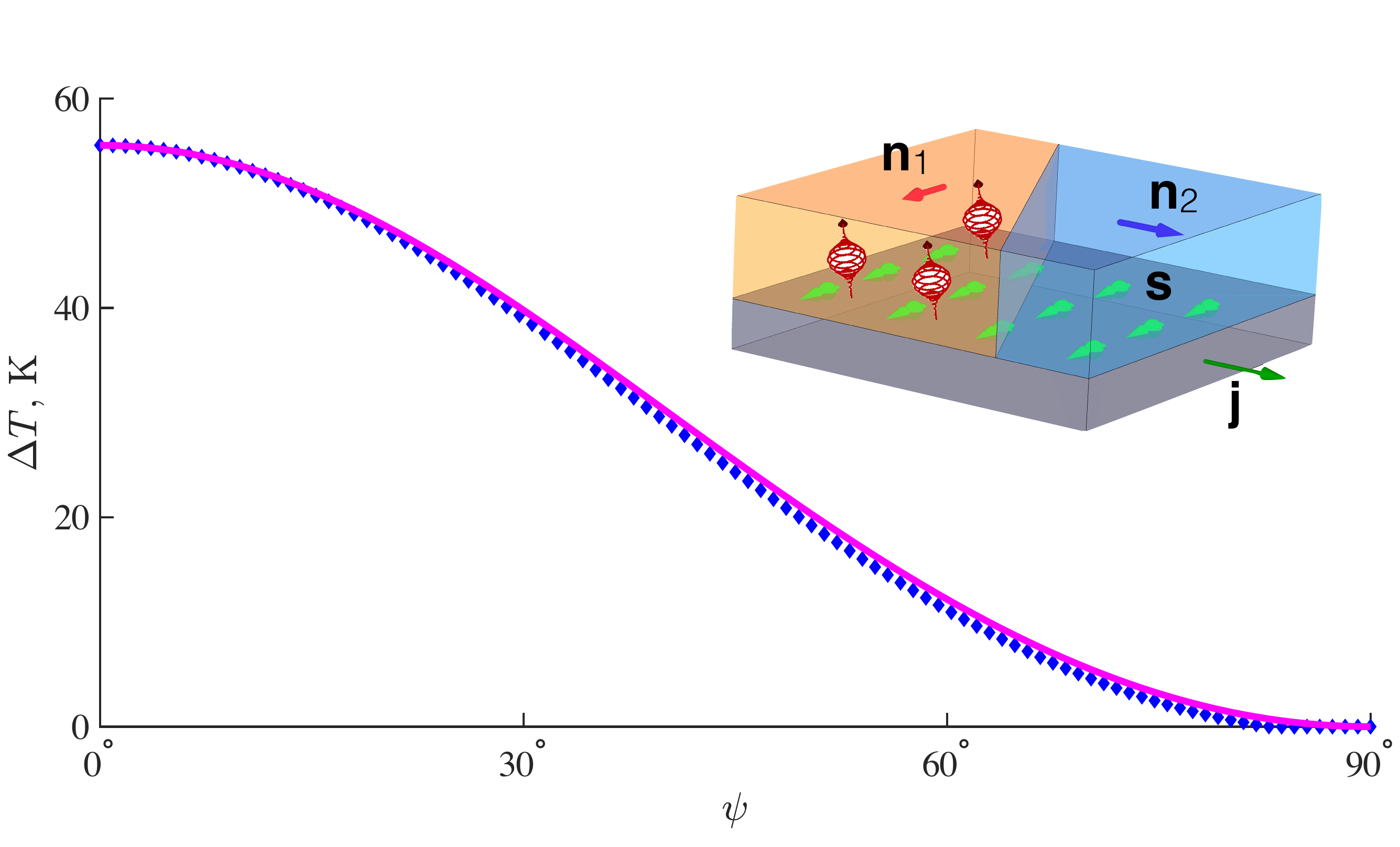}
	\caption{(Color online)  Angular dependence of the difference between the magnon and electron temperatures, $\Delta T$, calculated from Eq.~(\ref{eq_motion_antiferromagnet_initial}) (blue diamonds) at $H_\mathrm{dc}=1$~mT, $T=150$~K. Solid line shows approximation according to Eq.~(\ref{eq_magnon_gas}).  Inset: cartoon of the spin Peltier effect in an antiferromagnet. A charge current $\mathbf{j}$ through the heavy metal electrode (gray region in Inset)  creates a transverse spin current and generates a spin accumulation $\|\mathbf{s}$ at the interface.  In the left (orange) region, where the N\'eel vector $\mathbf{n}_1\|\mathbf{s}$ ($\psi=0$), the spin torque transfers angular momentum and energy from the electrons in the heavy metal to the magnons in an antiferromagnet thereby cooling the electrons and heating the magnons, effectively raising the magnon temperature $T_\mathrm{mag}$ with respect to the electron temperature $T$. In the right (blue) region, where the N\'eel vector $\mathbf{n}_2\perp\mathbf{s}$, there is no spin transfer from the electrons to the magnon gas and $T_\mathrm{mag}=T$. }
	\label{fig_Peltier}
\end{figure}

Thus, the temperature of the magnon gas is higher than $T$ in the antidamping-like region I of phase diagram Fig.~\ref{fig_phase_diagram}(a) and coincides with $T$ in the field-like region II. Fig.~\ref{fig_Peltier} shows angular dependence $\Delta T(\psi)$ calculated for KNiF$_3$ (blue diamonds). For the relatively small currents $H_\mathrm{dc}\le H^\mathrm{cr}_\mathrm{dc}$ this dependence is well approximated by $\cos^2\psi$ (magenta solid line) in the full range of angles.

From Eq.~(\ref{eq_magnon_gas}) it follows that the difference $\Delta T(\psi)$ is maximal when $\mathbf{s}\|\mathbf{n}^{(0)}$ and vanishes when $\mathbf{s}\perp\mathbf{n}^{(0)}$. This means, in particular, that the spin Peltier effect results in the effective temperature gradient between the 90$^\circ$ antiferromagnetic domains, as shown in inset in Fig.~\ref{fig_Peltier}. This gradient can create a force which moves the domain walls in antiferromagnets, similar to a force created  by "standard" temperature gradient \cite{Selzer2016, Kim2015d}.

Domain wall motion can be used to observe the spin Peltier effect in antiferromagnets. Really, geometry in which the external spin current induces domain wall motion via  the spin Peltier effect (inset in Fig.~\ref{fig_Peltier}) excludes direct spin orbit torque mechanism considered in Ref.~\cite{Shiino2016}. Thus, these two effects can be clearly separated in the experiment. Moreover, while temperature gradient creates the so-called ponderomotive force which pushes the domain wall toward the hotter region and eliminates unfavouralble domain, spin orbit torque induces simultaneous shifting of all the domain walls in the same direction, typical for the race-track memory \cite{Shiino2016}. This difference in behaviour can be also used to distinguish between two mechanisms which can be induced by the external spin current.

We also believe, that domain wall motion via the spin Peltier effect can find practical applications because i) in-plane geometry of the  external spin current is consisted with spin Hall effect; ii) the effective force, which acts on the domain wall, grows with temperature and can be much larger than that originating from spin transfer (orbit) torques.     
To confirm the second statement we compare the forces acting on the 90$^\circ$ domain wall between domains with $\mathbf{n}_1\|\hat{x}$ and $\mathbf{n}_2\|\hat{y}$ in two configurations: $\mathbf{s}\|\hat{x}$ (the force is induced by the spin Peltier effect) and $\mathbf{s}\|\hat{z}$ (the force is induced by the spin transfer torque). For this we use general expression for the forces acting on the domain wall 
(see Supplementary Materials in \cite{Gomonay2016}):
\begin{equation}\label{eq_force general}
F=\frac{1}{2\gamma^2M_sH_\mathrm{ex}}\int\mathbf{T}\cdot\mathbf{n}^{(0)}\times\partial_\xi\mathbf{n}^{(0)}d\xi,
\end{equation} 
where $\mathbf{T}$ are the torques in the r.h.s. of Eq.~(\ref{eq_motion_antiferromagnet_initial}), $\xi$ is the space coordinate in the direction of the domain wall normal, and ${\mathbf{n}^{(0)}}(\mathbf{r},t)$ satisfy equation of the domain wall motion in the absence of the external torques and damping.

The force generated directly by the spin transfer torque (in the simplest geometry where $\mathbf{s}\perp\mathbf{n}$ throughout the domain wall) is then $F_\mathrm{STT}=\pi H_\mathrm{dc}M_s$. The force created by the magnons due to the spin Peltier effect is $F_\mathrm{Peltier}=k_\mathrm{B}N\Delta T\approx k_\mathrm{B}NT(H_\mathrm{dc}/H^\mathrm{cr}_\mathrm{dc})^2$, where $N$ is the number of the excited magnons (for the details see Appendix). For KNiF$_3$ at $T=150$~K and $H_\mathrm{dc}=1$~mT the ratio $F_\mathrm{Peltier}/F_\mathrm{STT}\propto 100$.

The external spin current can also induce the spin Peltier effect in uniaxial antiferromagnets like Cr$_2$O$_3$ or MnF$_2$.  However, observation of the effect in these materials can be challenging due to the absence of the non-180$^\circ$ domains.

\section{Magnon (spin) Seebeck effect induced by the external spin current}
In the previous section we considered a simplified picture of the spin Peltier effect disregarding magnon dispersion. In this section we give more accurate description of spin caloritronic phenomena induced by the external spin current. We focus mainly on the Seebeck effect which is thermodynamically conjugated to the Peltier effect due to the Onsager reciprocal principle. 

The magnon Seebeck effect is defined as generation of the magnon flow with a current density $\mathbf{J}_\mathrm{mag}$ by temperature gradient $\boldsymbol{\nabla} T$. In contrast to ferromagnets, an antiferromagnetic magnon flux can be either non-polarized or can have a nonzero average spin (anti)parallel to $\mathbf{n}^{(0)}$. In the latter case we talk about the magnon \emph{spin} Seebeck effect in which the \emph{spin}  magnon current with the current density $\mathbf{J}^\mathrm{spin}_\mathrm{mag}$ is created. 

To illustrate the idea of the magnon (spin) Seebeck effect in antiferromagnets we use general
approach developed in Ref.~\cite{Rezende2016}. In assumption that the magnon spectra consist of two types of circularly polarized modes, each of which can carry "up" or "down" ($\uparrow$ or $\downarrow$) spins, the spin magnon current is calculated as a difference between the magnon fluxes with opposite spin direction. Ultimate expression for the spin magnon current density induced by the temperature gradient (disregarding spatial variation of the magnon accumulation) is:
\begin{equation}\label{eq_spin_magnon_current_expression}
\mathbf{J}^\mathrm{spin}_\mathrm{mag}=-\hbar\int \frac{d^3k}{(2\pi)^3}\left[\tau_\uparrow\frac{\partial f^{(0)}_\uparrow}{\partial T}-\tau_\downarrow\frac{\partial f^{(0)}_\downarrow}{\partial T}\right]\mathbf{v}^\mathrm{mag}(\mathbf{v}^\mathrm{mag}\cdot\boldsymbol{\nabla} T),
\end{equation}   
where $f^{(0)}_{\uparrow,\downarrow}$ are the equilibrium magnon density functions, $\tau_{\uparrow,\downarrow}$ and $\mathbf{v}^\mathrm{mag}_\uparrow=\mathbf{v}^\mathrm{mag}_\downarrow=\mathbf{v}^\mathrm{mag}$ are the magnon relaxation time and the magnon velocity, correspondingly, $\hbar$ is the Planck constant. All the magnon parameters ($f,\tau,\mathbf{v}^\mathrm{mag}$) are functions of the magnon wave vector $\mathbf{k}$.

Analysis of Eq.~(\ref{eq_spin_magnon_current_expression}) shows that no spin Seebeck effect can be observed as long as the spin-up and spin-down modes of an antiferromagnet are degenerate, so that $f^{(0)}_\uparrow=f^{(0)}_\downarrow$, $\tau_\uparrow=\tau_\downarrow$, and  $\mathbf{J}_\mathrm{mag}=0$. To remove degeneracy of the modes one can apply the external magnetic field $\mathbf{H}\|\mathbf{n}^{(0)}$, as it was done in the experiments \cite{Seki2015, Wu2016, Holanda2017}. The magnetic field splits the frequencies $\omega_{\uparrow,\downarrow}$ of the up and down modes ($\omega_{\uparrow}-\omega_{\downarrow}\propto H$) and has no influence on the relaxation times, $\tau_\uparrow=\tau_\downarrow$. In this case the spin Seebeck coefficient $S_\mathrm{See}$, which defines relation between the magnon spin current and temperature gradient, $\mathbf{J}_\mathrm{mag}=S^\mathrm{spin}_\mathrm{See}\boldsymbol{\nabla} T$ can be calculated from expression
\begin{equation}\label{eq_SSE_field}
S^\mathrm{spin}_\mathrm{See}(H)=-\hbar\int \frac{d^3k}{(2\pi)^3}\left[\frac{\partial f^{(0)}_\uparrow}{\partial T}-\frac{\partial f^{(0)}_\downarrow}{\partial T}\right]\tau\left(v^\mathrm{mag}\right)^2\propto H.
\end{equation}

\begin{figure}
	\centering
	\includegraphics[width=0.95\linewidth]{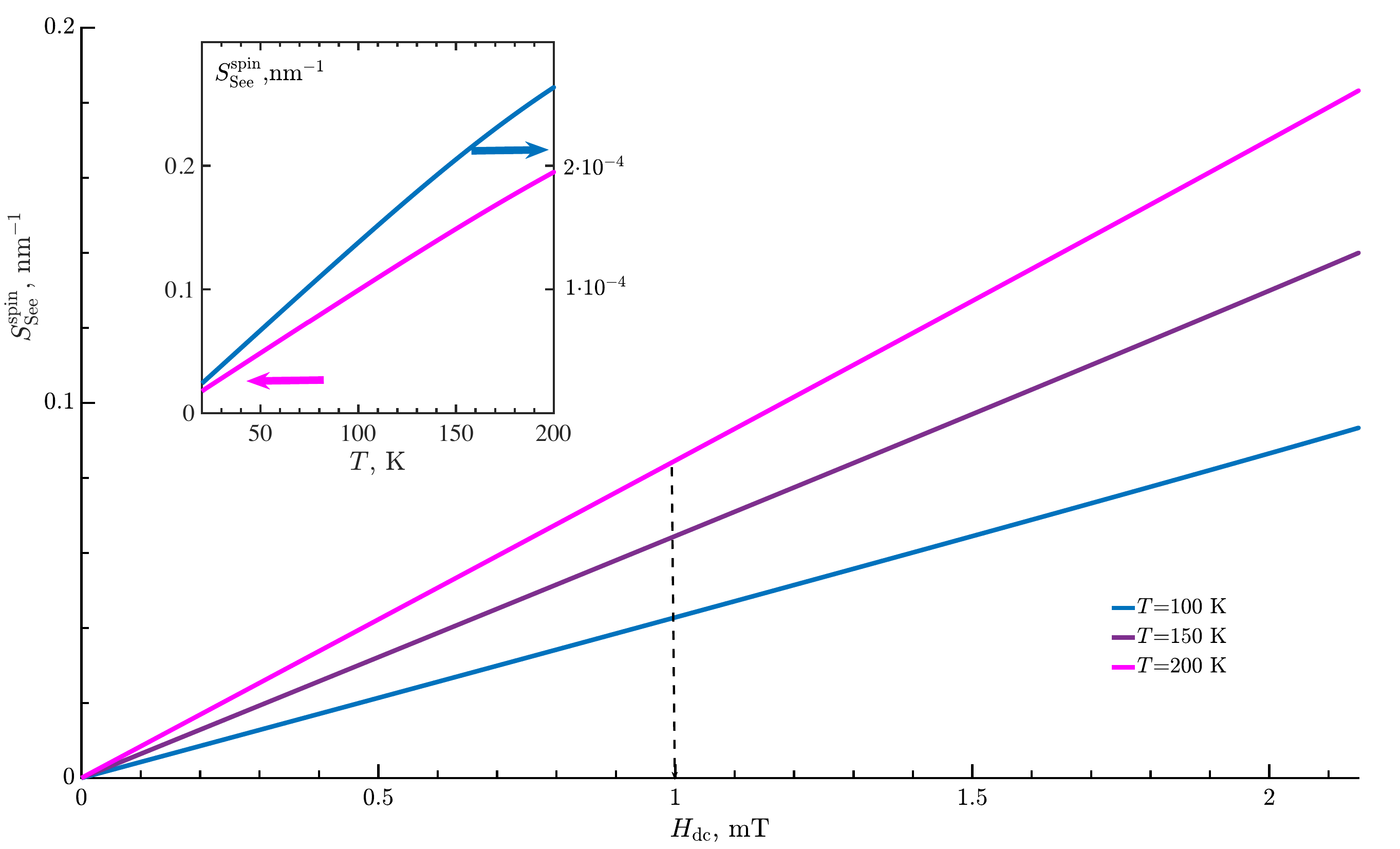}
	\caption{(Color online) Current dependence of the spin magnon Seebeck coefficient $S^\mathrm{spin}_\mathrm{See}$ for the different temperature values, $\mathbf{s}\|\mathbf{n}^{(0)}$. Inset: temperature dependence of $S^\mathrm{spin}_\mathrm{See}$ in the presence of the spin-polarized current (left axis, magenta) and the external magnetic field of the equivalent value (right axis, blue),  $H_\mathrm{dc}=$1~mT (vertical dashed arrow).}
	\label{fig_SEE_vs_current}
\end{figure}

Degeneracy of the magnon modes can be also removed by the external spin current, as it was discussed above. Assuming that the main contribution into relaxation comes from Gilbert damping, we find that  the relaxation times $\tau_{\uparrow,\downarrow}=1/\gamma^\mathrm{AF}_{1,2}$ and, according to Eq.~(\ref{eq_damping_parallel}),  $\tau_\uparrow-\tau_\downarrow\propto H_\mathrm{dc}(\mathbf{s}\cdot\mathbf{n}^{(0)})$, while the magnon frequencies are degenerate.  In this case $f^{(0)}_\uparrow=f^{(0)}_\downarrow$ and the spin Seebeck coefficient
\begin{equation}\label{eq_SSE_spin magnon current}
S^\mathrm{spin}_\mathrm{See}(H_\mathrm{dc}\mathbf{s})=-\hbar\int \frac{d^3k}{(2\pi)^3}\frac{\partial f^{(0)}}{\partial T}\left(\tau_\uparrow-\tau_\downarrow\right)\left(v^\mathrm{mag}\right)^2\propto H_\mathrm{dc}(\mathbf{s}\cdot\mathbf{n}^{(0)}).
\end{equation}  
Fig.~\ref{fig_SEE_vs_current} illustrates the field, spin current and temperature depencies of the spin Seebeck coefficient in KNiF$_3$ calculated from Eqs.~(\ref{eq_spin_magnon_current_expression}). For calculations we model the magnon dispersion as $\omega^\mathrm{AF}=\sqrt{\omega^2_\mathrm{AFR}+\gamma^2H\mathrm{ex}^2\sin^2(\pi ka)}$, where $a$ is the lattice constant, and assume Bose-Einstein distribution for $f^{(0)}_{\uparrow, \downarrow}$ (see, e.g. \cite{Rezende2016}). Calculations demonstrate linear dependence of $S^\mathrm{spin}_\mathrm{See}$ as a function of the external spin current, in consistence with the second equality in Eq.~(\ref{eq_SSE_spin magnon current}). The spin Seebeck coefficient also linearly grows with temperature, as a result of increasing  population of the thermal magnons. Our calculations also confirm linear field dependence of $S^\mathrm{spin}_\mathrm{See}(H)$, according to Eq.~(\ref{eq_SSE_field}). However, the absolute value of $S^\mathrm{spin}_\mathrm{See}(H)$ is three order of magnitude smaller than $S^\mathrm{spin}_\mathrm{See}(H_\mathrm{dc})$  (Fig.~\ref{fig_SEE_vs_current}, inset), that demonstrates high efficiency of the external spin current compared to the external magnetic field.

\begin{figure}
	\centering
	\includegraphics[width=0.95\linewidth]{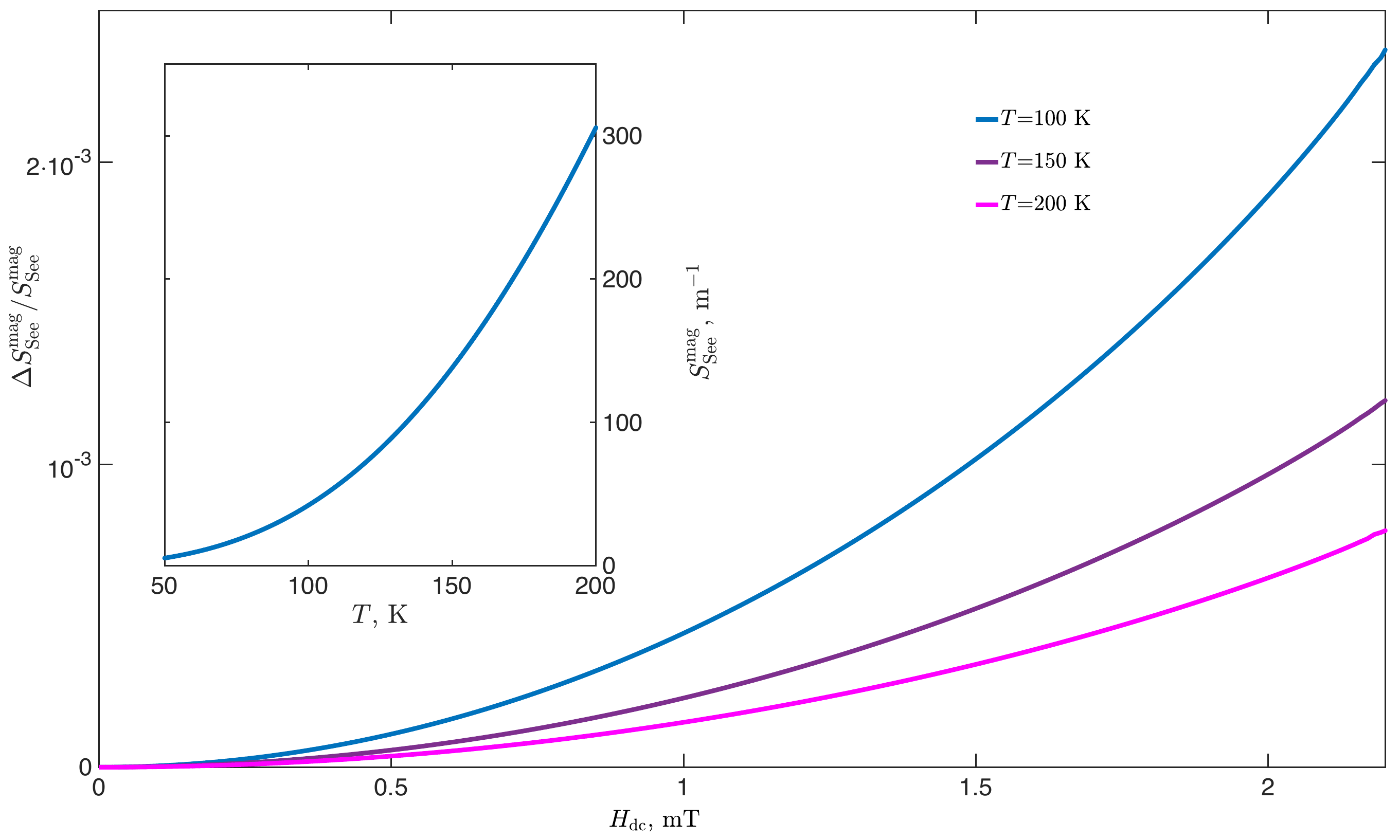}
	\caption{(Color online) Relative spin current-dependent contribution $\Delta S^\mathrm{mag}_\mathrm{See}/S^\mathrm{mag}_\mathrm{See}$ to the magnon Seebeck coefficient as a function of current for the different temperatures,  $\mathbf{s}\|\mathbf{n}^{(0)}$. Inset: temperature dependence of $S^\mathrm{mag}_\mathrm{See}$ at $H_\mathrm{dc}=$1~mT.}
	\label{fig_magnon_SE}
\end{figure}

While the spin magnon Seebeck effect is mediated by the circularly polarized magnons and thus can occur only under the certain conditions (e.g., in area I for cubic antiferromagnet, see Fig.~\ref{fig_phase_diagram}), the magnon Seebeck effect is not sensitive to the magnon polarization and ability to carry spin. This can be seen from the general expression for the magnon Seebeck coefficient    
\begin{equation}\label{eq_SSE_magnon current}
S^\mathrm{mag}_\mathrm{See}=-\hbar\int \frac{d^3k}{(2\pi)^3}\frac{\partial f^{(0)}}{\partial T}\left(\tau_\uparrow+\tau_\downarrow\right)\left(v^\mathrm{mag}\right)^2
\end{equation}  
which is nonzero even if $\tau_\uparrow=\tau_\downarrow$. However, in the antidamping region (area I) there is an additional contribution $\Delta S^\mathrm{mag}_\mathrm{See}\equiv S^\mathrm{mag}_\mathrm{See}(H_\mathrm{dc})-S^\mathrm{mag}_\mathrm{See}(0) \propto H_\mathrm{dc}^2$ (see Fig.~\ref{fig_magnon_SE}) which appears due to the energy transfer from electrons in heavy metal to the magnon gas, similar to the mechanism responsible for the spin Peltier effect. This contribution also grows with temperature  (inset in Fig.~\ref{fig_magnon_SE}), similar to $S^\mathrm{spin}_\mathrm{See}$.  

\section{Conclusions}
To summarize, we considered spin caloric effects in antiferromagnets induced by the external spin polarized current. These effects steam from the spin exchange between the spin-polarized electrons and thermal magnons and thus are demanding to ability of magnons to transfer spin. This means, in particular, that all the considered effects should be pronounced in antiferromagnets whose spin-wave eigen modes are circularly polarized, i.e. in easy-axis or cubic materials. In antiferromagnets with easy-plane type of the magnetic ordering spin transfer between electrons and magnons is hindered due to the incoherence of the linearly-polarized modes. In this case interpretation of spin caloric effects needs more sophisticated models. 

We demonstrate that the spin caloric effects assisted by the external spin current can be much more effective than the spin caloric effects in the presence of the external magnetic field (spin magnon Seebeck effect) or pure "mechanical" effects of spin current (spin Peltier effect). The predicted spin Peltier effect in combination with the temperature gradient can be used 
for the precise manipulation of the 90$^\circ$ domain walls. The (spin) magnon Seebeck effect should be taken into account in the experiments with spin Hall geometry, as in this case the spin-polarized current can create additional magnon diffusion. 
\acknowledgements
This work was supported by the DFG-SPP SpinCat Grant, the Alexander von Humboldt Foundation,  the DFG (project SHARP 397322108), the EU FET Open RIA Grant no. 766566 and the Transregional Collaborative Research Center (SFB/TRR) 173 SPIN+X. O.G. also acknowledge fruitful discussions with Profs. Ya. Tserkovnyak and M. Kl\"aui.
\bibliographystyle{iopart-num}

\providecommand{\newblock}{}

\appendix
\section{Motive forces induced by spin current and fluctuations}
In this section	we anaylise contribution of the  magnon fluctuations into the thermomagnonic torques and calculate the motive forces which act on the domain wall in the presence of nonequilibrium magnon distribution between two domains with the equilibrium N\'eel vector $\mathbf{n}^{(0)}_{1,2}$ (see, e.g. Fig.\ref{fig_Peltier}, inset).

Following the procedure suggested in Ref.\cite{Kim2015c} for analysis of thermomagnonique torques in ferromagnets, we split the unit N\'eel vector $\hat{\mathbf{n}}$ into two orthogonal components:
\begin{equation}\label{eq_unit_Neel_vector}
\hat{\mathbf{n}}=\sqrt{1-(\delta\hat{\mathbf{n}}^2)}\hat{\mathbf{n}}^{(0)}+\delta \hat{\mathbf{n}}.
\end{equation}
The slowly varying component $\hat{\mathbf{n}}^{(0)}(\mathbf{r},t)$ describes the equilibrium antiferromagnetic domain wall. The fast component $\delta\hat{\mathbf{n}}$ describes fluctuations. 

Substituting relaltion (\ref{eq_unit_Neel_vector}) into Eq.(\ref{eq_motion_antiferromagnet_initial}) we get equation of motion for the N\'eel vector $\hat{\mathbf{n}}^{(0)}(\mathbf{r},t)$  with fluctuations:
\begin{subequations}\label{eq_domain_wall}
	\begin{align}
	\hat{\mathbf{n}}^{(0)}\times(\ddot{\hat{\mathbf{n}}}^{(0)}-\gamma^2H_\mathrm{ex}\mathbf{H}_\mathbf{n}^{(0)})=&-2\alpha_\mathrm{G}\gamma H_\mathrm{ex}\hat{\mathbf{n}}^{(0)}\times\dot{\hat{\mathbf{n}}}^{(0)}\\
	&+\gamma^2 H_\mathrm{ex}H_\mathrm{dc}\hat{\mathbf{n}}^{(0)}\times\mathbf{s}\times\hat{\mathbf{n}}^{(0)}\label{eq_domain_wall_1}\\
	&-\omega_\mathrm{AFR}^2\hat{\mathbf{n}}^{(0)}\times\frac{\partial}{\partial \hat{\mathbf{n}}}\left(\left.\Lambda_{jk} \right|_0\langle\delta \hat{n}_j\delta \hat{n}_k\rangle\right)\label{eq_domain_wall_2}\\
	&+c^2\langle\delta\hat{\mathbf{n}}\times\nabla^2 \delta{\hat{\mathbf{n}}}\rangle-c^2\hat{\mathbf{n}}^{(0)}\times\partial_j\hat{\mathbf{n}}^{(0)}\partial_j\langle\delta\hat{\mathbf{n}}^2\rangle\label{eq_domain_wall_3}\\
	&+\frac{\partial}{\partial t}\langle\delta\hat{\mathbf{n}}\times \delta\dot{\hat{\mathbf{n}}}\rangle-\omega_\mathrm{AFR}^2\langle\delta\hat{\mathbf{n}}\times\hat{\Lambda}\delta\hat{\mathbf{n}}\rangle\label{eq_domain_wall_4},
	\end{align}
\end{subequations}
where brackets $\langle\ldots\rangle$ mean statistical average. 

The terms (\ref{eq_domain_wall_1})-(\ref{eq_domain_wall_4}) in the r.h.s. of Eq.~(\ref{eq_domain_wall}) are the torques $\mathbf{T}$ which can contribute into the effective force acting on the domain wall. According to Eq.~(\ref{eq_force general}), the force is calculated as an integral
\begin{equation}\label{eq_force_particular}
F=\frac{2M_s}{\gamma^2H_\mathrm{ex}}\int\mathbf{T}\cdot\hat{\mathbf{n}}^{(0)}\times\partial_\xi\hat{\mathbf{n}}^{(0)}d\xi,
\end{equation} 
where $\hat{\mathbf{n}}^{(0)}(\mathbf{r},t)$ satisfy Eq.~(\ref{eq_domain_wall}) in the absence of the external torques and damping.

We consider separately contribution of each of the torques  (\ref{eq_domain_wall_1})-(\ref{eq_domain_wall_4}).

The torque (\ref{eq_domain_wall_1}) origination from a spin current creates a force 
\begin{equation}\label{eq_STT_torque}
F_\mathrm{STT}={2\gamma M_s}H_\mathrm{dc} \int(\mathbf{s}\times\hat{\mathbf{n}}^{(0)})\cdot\partial_\xi\hat{\mathbf{n}}^{(0)}d\xi.
\end{equation}
This force is zero if $\mathbf{s}$ lyes within the plane, in which the N\'eel vector rotates from $\mathbf{n}_1$ to  $\mathbf{n}_2$. If $\mathbf{s}$ is perpendicular to this plane, for a 90$^\circ$ domain wall one gets from (\ref{eq_STT_torque}) that $F_\mathrm{STT}={\pi\gamma M_s}H_\mathrm{dc}$.

The fluctuation-dependnet torque (\ref{eq_domain_wall_2}) describes the magnon contribution into anisotropy energy. This torque creates a Peltier force
\begin{equation}\label{eq_Peltier_Force}
F_\mathrm{Peltier}=\frac{2M_s}{\gamma^2H_\mathrm{ex}}	\omega^2_\mathrm{AFR}\left(\left.\left.\Lambda_{jk} \right|_0 \langle\delta \hat{n}_j\delta \hat{n}_k\rangle\right|_{\mathbf{n}_1}-\left.\left.\Lambda_{jk} \right|_0 \langle\delta \hat{n}_j\delta \hat{n}_k\rangle\right|_{\mathbf{n}_2}\right).
\end{equation}

To calculate correlators $\langle\delta \hat{n}_j\delta \hat{n}_k\rangle$ we treat Eqs.~(\ref{eq_motion_antiferromagnet_initial}) and (\ref{eq-linear_excitations})  as  a set of independent Langevin equations for each of the AF eigen modes $\delta\hat{\mathbf{n}}_{1,2}$:
\begin{equation}\label{eq-Langevin}
\delta\ddot{\hat{\mathbf{n}}}_{1,2}+2\gamma^\mathrm{AF}_{1,2} \delta\dot{\hat{\mathbf{n}}}_{1,2}+\left(\omega^\mathrm{AF}_{1,2}\right)^2\delta\hat{\mathbf{n}}_{1,2}=\gamma^2 H_\mathrm{ex}\mathbf{h}_{\mathbf{n}1,2},
\end{equation}
where the damping coefficients $\gamma^\mathrm{AF}_{1,2}$ and eigen frequencies $\omega^\mathrm{AF}_{1,2}$, defined in the main text, are the functions of local equilibrium orientation of the N\'eel vector $\hat{\mathbf{n}}^{(0)}(\mathbf{r})$.  For each of the eigen mode
\begin{equation}\label{eq_fluctuation_correlator}
\langle\delta \hat{n}^2_{1,2}\rangle=\frac{(\gamma^2H_\mathrm{ex})^2D}{\pi\gamma^\mathrm{AF}_{1,2}\left(\omega^\mathrm{AF}_{1,2}\right)^2}
\end{equation} 
Substituting relation (\ref{eq_diffusivity}) into (\ref{eq_fluctuation_correlator}) in approximation of high temperature $\hbar\omega_\mathrm{AFR}\ll k_\mathrm{B}T$ we get 
\begin{equation}\label{eq_correlation}
\left.	\omega^2_\mathrm{AFR}\Lambda_{jk} \right|_0 \langle\delta \hat{n}_j\delta \hat{n}_k\rangle=\gamma^2H_\mathrm{ex}k_\mathrm{B}(T_1^\mathrm{eff}+T_2^\mathrm{eff}).
\end{equation}
Corresponding contribution into Peltier force is then $F_\mathrm{Peltier}=k_\mathrm{B}N\Delta T$, where $N$ is the number of the excited magnon modes.

The torques (\ref{eq_domain_wall_3}) describe so-called thermomagnonic torques also predicted in ferromagnets \cite{Kim2015c}. The first term describes spin transfer torque which occurs due to spin polarization of the magnon flux. As in our case spin polarization (magnetization) of the eigen modes is parallel to $\mathbf{n}^{(0)}$ (see Eq.~(\ref{eq_mode_magnetization})), this term does not contribute into the motive force. The second term is significant in the presence of the external fields and/or a temperatue graduent which splits the magnon population in right and left domains. Its contribution into the motive force is similar to $F_\mathrm{Peltier}$.

The torques (\ref{eq_domain_wall_4}) are related with fluctuations of magnetization and does not contribute into the motive force, as magnetization is parallel to $\mathbf{n}^{(0)}$ .

%
%
%

\end{document}